\documentclass[12pt]{article}
\usepackage{amsmath,euscript,amssymb}
\usepackage[dvips]{graphicx}
\def\v1{\vspace{1cm}}
\def\be{\begin{equation}}
\def\ee{\end{equation}}
\def\bc{\begin{center}}
\def\ec{\end{center}}

\newcommand{\bea}{\begin{eqnarray}}
\newcommand{\eea}{\end{eqnarray}}

\newcommand\ba{\begin{eqnarray}}
\newcommand\ea{\end{eqnarray}}

\newcommand{\br}[1]{\left( #1 \right)}

\newcommand{\brm}[1]{\left| #1 \right|}

\newcommand{\GeV}{~\mbox{GeV}}
\newcommand{\MeV}{~\mbox{MeV}}

\begin{document}

\title{
   Pion Polarizability in the NJL model
and Possibilities of its Experimental Studies
in Coulomb Nuclear Scattering
}
\author{Yu. M. Bystritskiy, A.V. Guskov, 
V.N. Pervushin, M.K. Volkov \\[0.3cm]
{\normalsize\it Joint Institute for Nuclear Research},\\
 {\normalsize\it 141980, Dubna, Russia.} }
\date{\empty}
\maketitle
\medskip

\begin{abstract}
   The charge pion polarizability  is calculated in
the  Nambu-Jona-Lasinio model,
 where  the quark loops (in the  mean field  approximation) and the meson loops
 (in the $1/N_c$ approximation)
   are taken into account.
   We show that quark loop contribution dominates, because
  the  meson loops  strongly conceal each other.
 The sigma-pole contribution $(m^2_\sigma-t)^{-1}$  plays the main role
 and contains
strong t-dependence of the effective pion polarizability
at the region
$|t|\geq 4M_\pi^2$.
 Possibilities of  experimental test of this  sigma-pole effect
 in the  reaction of Coulomb Nuclear Scattering
     are estimated for the COMPASS experiment. 
\end{abstract}
\section{Introduction}

Elementary particle polarizabilities $\alpha^E,~\alpha^H$ were introduced
as coefficients of low energy expansion of the Compton effect
amplitude \cite{Klein:1955zz}
using the definition of the
effective potential energy:
\ba\label{p-1}
V_{\rm eff}=-\frac{1}{2}[\alpha^E E^2+\alpha^H H^2]\Big|_{\alpha^E=-\alpha^H}=\frac{\alpha^E}{4}F^2_{\mu\nu}.
\ea
 These coefficients
  as well as the electromagnetic radius are
constants characterizing the internal  structure of particles.

The values of charged pion polarizabilities $\alpha^H,\alpha^{\rm E}_\pi$  was measured
in Serpukhov \cite{Serp1}
\ba\label{p-1e}
\alpha_\pi^E\simeq-\alpha^{\rm H}_\pi = (6.8 \pm 1.8)\cdot 10^{-4} {\mbox{\rm fm}}^{3}
\ea
 and  MAMI \cite{9e}
\ba\label{p-1ee}\alpha_\pi^E\simeq-\alpha^{\rm H}_\pi = (5.8 \pm 1.7)\cdot 10^{-4} {\mbox{\rm fm}}^{3}
\ea
 and was extracted from the MARK II data \cite{MarkIICollab} in \cite{10e}
\ba\label{p-1eee}\alpha_\pi^E\simeq-\alpha^{\rm H}_\pi = (2.2\pm1.6) \cdot 10^{-4} {\mbox{\rm fm}}^{3}.\ea

One can see that the precision of the  experimental measurements
is too  low to distinguish between the many predictions
of the value of the
charged pion polarizability
obtained in various quark,
chiral, dispersion and other models (see e.g. the reviews \cite{1979,Review,Volkov:1986zb,2006,gasser}).

There is a hope that new more precise measurements of the pion polarizabilities
at the COMPASS experiment at CERN \cite{compass,Moinester,moin,sasha}
provide a good opportunity for the verification of these models.

The idea to investigate the charged pion polarizability in radiative $\pi^-$ meson
scattering in the nuclear Coulomb field was proposed in \cite{Idea}.
It was shown in \cite{Idea} that in the reaction
\ba
    \pi^-+(A,Z)=\pi^-+(A,Z)+\gamma,
\ea
the Coulomb amplitude dominates for very small four-momentum
transfers $\brm{t} \leq 2\times 10^{-4} \br{\GeV/c}^2$
and the contribution from the pion polarizability to the Compton
effect increases with the decrease of the Coulomb transfer.

The first experiment proposed in \cite{Idea}
was fulfilled at SIGMA-AYAKS spectrometer \cite{Serp1} in the context of the first predictions
 of the pion polarizability value in
the \textit{quantum field theory} approach \cite{1979,Kazakov} to  the
non-polynomial \textit{Effective Chiral  Lagrangian}
\cite{1967}.
The results of calculation in \cite{75}
 can be presented as the sum
 of both the fermion loops  and the meson ones 
\ba\label{124z}
  {\alpha^E_{\pi^\pm}}(t)&=&
  {\br{\alpha_{\pi^\pm}}_{ch}}
    \left[{\beta^{\rm fermion}_{\pi^\pm}}(t)+{\beta^{\rm pion}_{\pi^\pm}}(t)\right],
    \ea
 where $t$ is given in the experimental region $|t/(2M_\pi)^2|\sim 1$ \cite{Serp1} and
 \ba\label{124}
{\br{\alpha_{\pi^\pm}}_{ch}}= \dfrac{\alpha}{2\pi F^2_\pi m_\pi}
= 5.8\times 10^{-4}fm{}^{-3};
\ea
is the chiral limit, here $F_\pi=93$ MeV and  $\alpha=\dfrac{e^2}{4\pi}=\dfrac{1}{137}$.
The baryon loops gave the main contribution ${\beta^{\rm fermion}}_{\pi^{\pm}}\simeq 1$,
whereas the meson loop
contribution was small and negative ${\beta^{\rm pion}}_{\pi^\pm} \simeq  -0.1$.
It disagrees with the value ${\beta^{\rm pion}}_{\pi^\pm}\simeq +0.5$
obtained in \cite{2009} in the Chiral Perturbation Theory  \cite{gl-1984}.
The authors of \cite{2009} associated
this value ${\beta^{\rm pion}}_{\pi^\pm}\simeq +0.5$  with
additional Chiral Lagrangians at order $p^4$. These low energy Lagrangians
can contain the  hadron contributions  including the fermion loop one
 in the hidden form.

  The problem of the ambiguities  of the pion polarizability obtained from the
  Effective Chiral Lagrangians
  can be clear up by both the direct experimental measurements and  the results obtained on
   the
  fundamental level of the QCD motivated quark models.

The calculations of the pion polarizability
in  the   quark NJL model \cite{Volkov:1986zb,1982,1983} motivated
by QCD  \cite{1989,1994} were fulfilled
 in \cite{Volkov:1986zb,1985} in the framework
 of the  mean field  approximation,
 where the quark loops only  are taken into account
 and the result ${\beta^{\rm quark}}^{\rm NJL}_{\pi^{\pm}}\simeq 1$ was obtained
 in agreement with the quark-baryon duality \cite{1981}.

In this paper, we take into account
 also the meson loops.
 They appear  in the next over $(1/N_c)$ approximation.
  The NJL model results on the charge pion polarizability are summed and compared
 with other  theoretical models.

 The possibilities of the experimental tests of the NJL model predictions
in reaction of Coulomb Nuclear Scattering
   will be estimated for the COMPASS experiment.

 The content of the paper is the following.
In Section 2.,
the kinematics of  experiments \cite{Serp1} was chosen to detect with good efficiency the
Compton effect events on pion with photon energies in the range $70-900\MeV$
in the pion rest frame. The pion polarizability is calculated in Section 3.
 Section 4 is devoted to discussion of
 possibility of the experimental test of the prediction of NJL model at COMPASS.

\section{Kinematics of Pion Compton Effect}

We consider the process: $\pi^-+(A,Z)=\pi^-+(A,Z)+\gamma$
\bea
\pi[p_1]~~+~~\gamma^{*} [q_1]~~\to~~
\pi[p_2] ~~+~~
\gamma [q_2],
\eea
where the  components of 4-vectors  $[p_1, q_1, p_2, q_2]$ are chosen in the form
\bea
&&p_1=\left(\varepsilon,~~~~~~~~~~~~~\varepsilon-\frac{M^2_\pi}{2\varepsilon},~~~~~~~~~~~~~~~~~~~0,~~~~0\right)\\
&&q_1=\left(\frac{Q^2}{2M_{\rm Nuc}},~~~~~{Q}=-\frac{M^2_\pi\omega^2+p_t^2}{2\omega(1-\omega)\varepsilon},
~~~~~~0,~~~~0\right)\\
&&p_2=\left((1-\omega)\varepsilon,~~
(1-\omega)\varepsilon-\frac{M^2_\pi+p_t^2}{2(1-\omega)\varepsilon},~~~p_t,~~~~0\right) \\
&&q_2 =\left(\varepsilon \omega,  ~~~~~~~~~~~\varepsilon \omega-\frac{p_t^2}{2\omega\varepsilon},
~~~~~~~~~~~~~-p_t,~~~~0\right).
\eea
 where $\varepsilon$ is the energy of incoming pion,  $M_{\rm Nuc}$ is the mass of the nuclear target and $\omega$ is the relative energy of emitted photon.
\bea
Q=-\frac{\omega^2 M_\pi^2+p^2_{t}}{2\varepsilon \omega(1-\omega)}\equiv
\frac{t}{2\varepsilon\omega},~~~~~~~~
\eea
is four-momentum transfer, and
\bea\label{t-1}
t=(p_1-p_2)^2=-\frac{ M_\pi^2\omega^2+p^2_{t}}{(1-\omega)}
\eea
is one of Mandelstamm variables.

Amplitude is
$$
A=A_ce^{i\phi}+A_s,
$$
where
$$
A_c=(4\pi)^{3/2}e^3\frac{4M_NZ\varepsilon^{\mu}}{Q^2}
\left\{g_{0\mu} - \frac{\varepsilon p_{2\mu}}{(p_2q_2)}+
\frac{(1-\varepsilon) p_{1\mu}}{(p_1q_2)} +\beta[g_{0\mu}(q_2q_1)-\varepsilon\omega q_{1\mu}]\right\}
$$
is the Coulomb amplitude; $A_s=(4\pi)^{3/2}e^2M_N\varepsilon^{\mu}T_{\mu}$ is
amplitude of the nuclear scattering; $\phi$ is the phase of
the Coulomb - nuclear scattering; Z is the charge of a nucleus;
$M_N$ is the mass of a nucleus; $\varepsilon^{\mu}$ is the vector of
the polarization of a photon; $g_{\mu\nu}={\rm diag(1,-1,-1,-1)}$;
 $\alpha_\pi=\alpha\beta_{-}/ M_\pi$ is the polarizability of a  pion $\pi^-$;
$T_{\mu}$ is the amplitude of a nuclear radiation scattering.

\section{Polarizability of a pion in NJL model}

 \begin{figure}[t]
 \begin{center}
 \includegraphics[width=0.65\textwidth,clip]{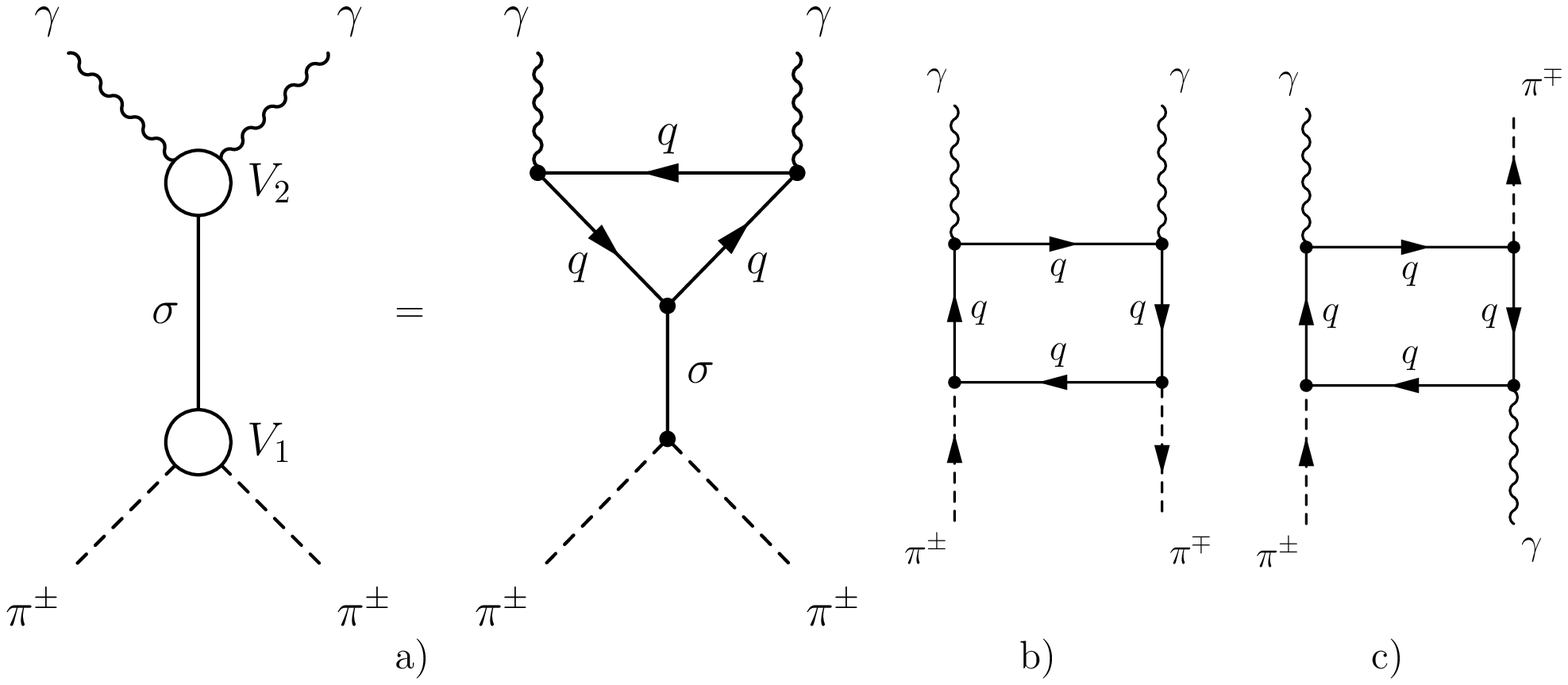}
\caption{
\label{fig2c}
}
\end{center}
\end{figure}

\subsection{Mean field approximation}

The calculations of the pion polarizability
in  the    NJL model \cite{Volkov:1986zb,1982,1983}  were fulfilled
 in \cite{Volkov:1986zb,1985} in the   approximation.
 It was shown in \cite{Volkov:1986zb,1985} that in this mean field approximation
 the main contribution
 goes from the
  the quark loops given in Figs. 1a,b,c. The first  diagram in Fig. 1a contains two
  vertexes\footnote{The functional dependence of
    the triangle diagram (in Fig. 1a) is determined by  the
      fermion loop integral \cite{Ebert:1996pc}. 
      The analytical form of this integral is
 \ba\label{12}
   \overline{F}(t_u) &=&\frac{3}{2}\left[1-\frac{{\Phi}(t_u)}{3}\right]=1+...,\\\label{13}
   {\Phi}(t_u)&=&-\frac{1}{t_u}\left[ 1+{t_u}^{-1}\phi(t_u)\right]=1 +...
\\\label{14}
    \phi(t_f)&=&\frac{1}{4}\ln^2\frac{\sqrt{1-1/t_u}+1}{\sqrt{1-1/t_u}-1}=\ln^2[{\sqrt{1-t_u}+\sqrt{-t_u}}]=
    -t_u - t_u^2...
\ea
  }
  \bea
  V_1=\dfrac{4m_u^2}{F_\pi}\sqrt{Z},~~
  V_2=\dfrac{10 \alpha}{9\pi F_\pi\sqrt{Z}}\overline{F}(t_u) 
  [g_{\mu\nu} q_1\cdot q_2-{q_1}_\mu {q_2}_\nu]\varepsilon_\mu(q_1)\varepsilon_\nu(q_2)
  \eea
  and the sigma meson propagator $1/(m_\sigma^2-t)$, where $m_u=280$ MeV is the constituent mass of the
   u-(d-) quarks and $t_u = t/(4 m_u^2)< 0$
  is  given in the region of negative values, 
  the factor $Z$ has a form $Z = \left(1-\frac{6m_u^2}{M_{A_1}^2}\right)^{-1}$;
  here $M_{A_1}=1260 MeV$ is the mass of $A_1$-meson \cite{Amsler:2008zzb}.

   The t-dependence of   the radiative triangle and box diagrams
   is very weak  
   and it can be neglected.
 In particular,
 the triangle diagram formfactor in the region of measurement $|t_{u}|\sim 1/4$
can be identified with the unit 1
\ba\label{11}
    \overline{F}(t_u) &=& 3\int\limits_{0}^{1} dx \int\limits_{0}^{1} dy y\frac{1-4y^2x(1-x)}
    {1-4y^2x(1-x)t_u}=1+t_u \frac{7}{90}...\\
    \ea
 within order of  2\% accuracy.
 %

 Taking into account the contribution
 of the diagrams in Figs. 1b,c in the lowest $q_1\cdot q_2$ approximation
 \bea
 \frac{\alpha}{\pi F_\pi}
  [g_{\mu\nu} q_1\cdot q_2-{q_1}_\mu {q_2}_\nu]\varepsilon_\mu(q_1)\varepsilon_\nu(q_2)
  \eea
 one can obtain the next expression for polarizability \cite{1985}
  \ba
  {\alpha^E_{\pi^\pm}}(t)= \dfrac{\alpha}{18\pi F^2_\pi m_\pi}
  \left[\frac{40m^2_u}{m_\sigma^2-t}
     - {1}\right]={\br{\alpha_{\pi^\pm}}_{ch}}
    {\beta^{\rm quark}_{\pi^\pm}}(t),\ea
    where the $t$-dependence  of the quark loops is neglected.


Using the definitions (\ref{124z}) and (\ref{124}) and the NJL relation \cite{Volkov:1986zb}
\ba\label{123}
m^2_\sigma=4m_u^2+M^2_\pi,
\ea
one can obtain the final result  in the form
 \ba \label{p-2}  \beta^{\rm quark}_{\pi^\pm}(t)=\frac{10}{9}\frac{4m^2_u}{m_\sigma^2-t}
     - \frac{1}{9}
     \ea
where the chiral limit
($t=0$, $M_\pi^2 = 0$ and $m^2_\sigma = 4 m_u^2$) corresponds to  unit
$\beta^{\rm quark}_{\pi^\pm}(0) \to 1$.

\subsection{Meson loops}

 \begin{figure}
\begin{center}
\includegraphics[width=20pc]{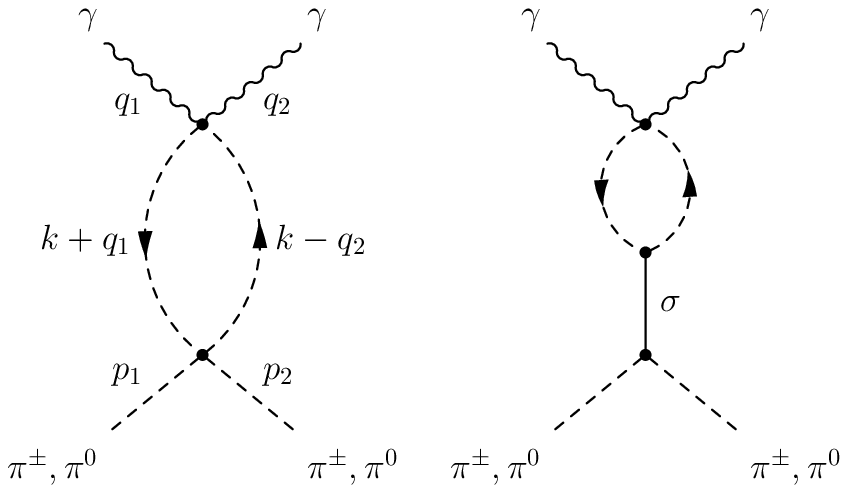}
\caption{
}
\end{center}
\end{figure}
\begin{figure}
\begin{center}
\includegraphics[width=20pc]{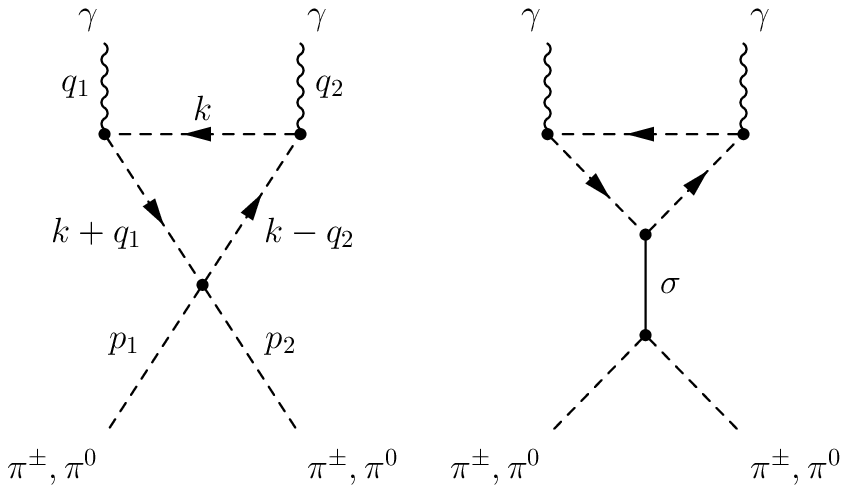}
\caption{
}
\end{center}
\end{figure}
\begin{figure}[h]
\begin{center}
\includegraphics[width=30pc]{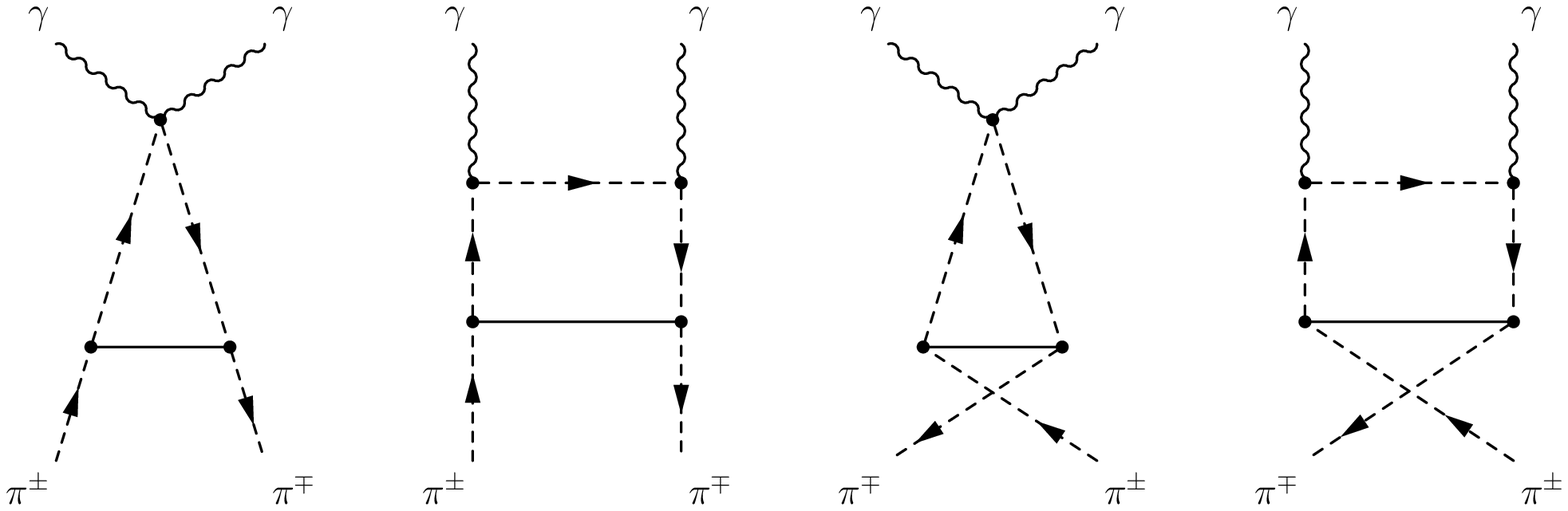}
\caption{
}
\end{center}
\end{figure}

 \begin{figure}
 \begin{center}
 \includegraphics[width=0.75\textwidth,clip]{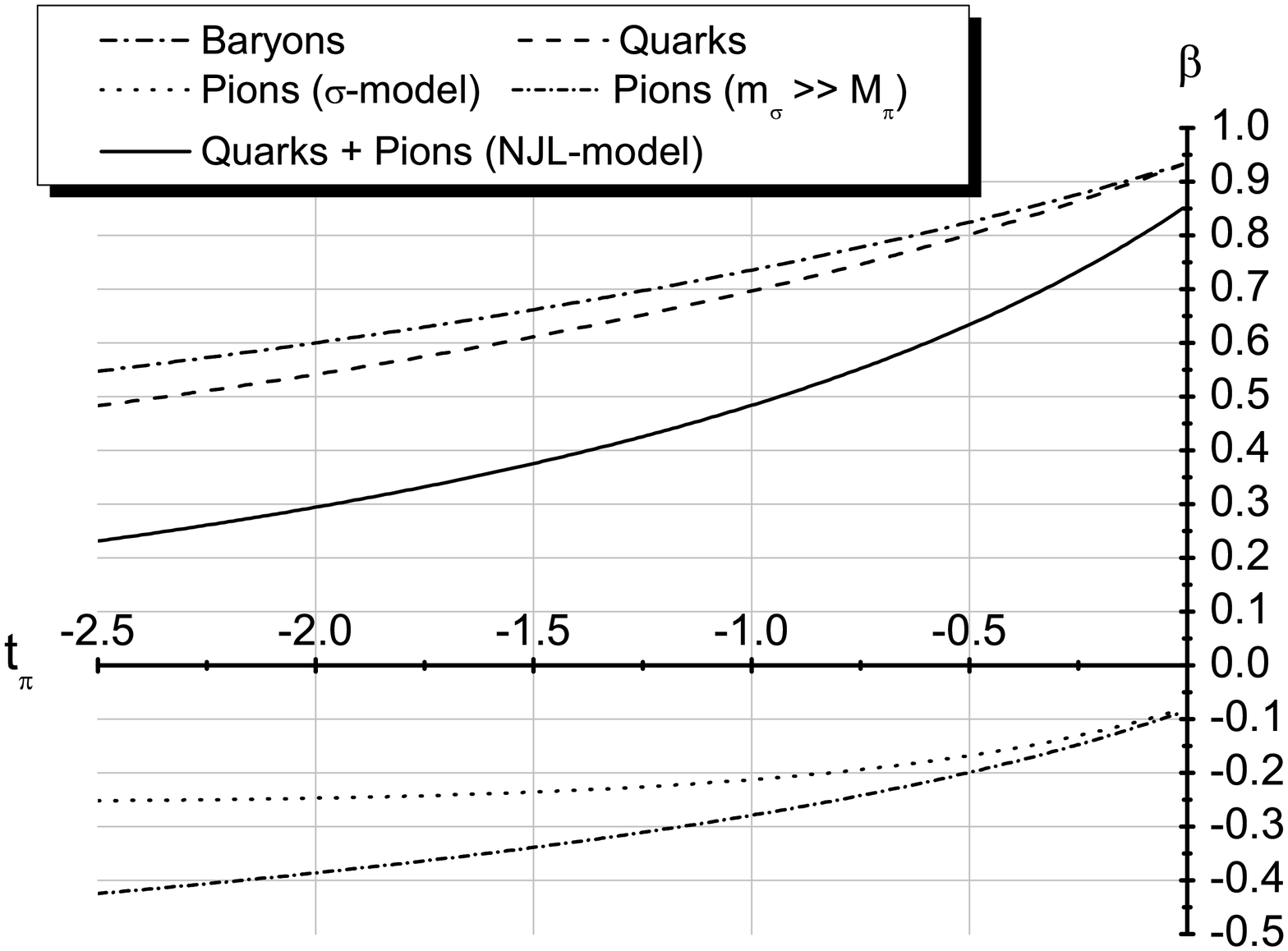}
\caption{\small These are the pion polarizability
contributions (ordinate axis) taken separately
and  their sum (quarks + mesons) given by Eq. (\ref{tot-1})
in terms of
$t_\pi=t/(2M_\pi)^2$ (abscissa). 
\label{fig2b}
}
\end{center}
\end{figure}

As it was shown in \cite{Volkov:2009mz}
in two-photon decays of scalar mesons
besides the quark loops contribution in the mean field approximation
the important role is played by meson loops in the next $1/N_c$ order approximation.
In particular, in two-photon decays of $f_0(980)$ meson, the meson loops plays
the dominant role \cite{Volkov:2009mz,Bystritskiy:2007wq}.
The comparably large  values of the meson loop contributions in comparison with the quark loops
caused by the fractional electric charge of quarks, while the mesons have an integer charges.
Therefore, in description of radiative decays of scalar meson
both the  quark and meson loops  are necessary to take them into account.
 In papers \cite{Volkov:2009mz} it was shown that
this approach leads to satisfactory agreement with the recent experimental data on
two-photon decays of scalar mesons $\sigma(600)$, $f_0(980)$ and $a_0(980)$.

Therefore, in description of pion polarizability in the electromagnetic vertex
$\sigma\to 2\gamma$ the quark triangle loop should be supplied by meson loop contributions.

However, in the case of the Compton effect, there is a set of diagrams with
internal sigma meson line. These diagrams also give the noticeable contribution
to polarizability and have an opposite sign in comparison with the meson diagrams on Fig. 4.

As a result the contributions of the meson loops in Figs. 2,3
strongly conceal the ones of meson loops in Fig. 4.

Finally the pion loop contributions take the form
\ba
\beta^{\rm pion}_{\pi^\pm}(t) =\frac{m_u^2}{M_\pi^2}
\left[\frac{4m_u^2}{m^2_\sigma-t}-1\right]
 \frac{\Phi(t_\pi)}{3}=\frac{m^2_\sigma-M_\pi^2}{m^2_\sigma-t}
 \left(t_\pi -\frac{1}{4}\right)\frac{\Phi(t_\pi)}{3},
\ea
where in this model $m^2_\sigma=4m_u^2+M_\pi^2$ and ${\Phi}(t_\pi)$  is the function of
$t_\pi=t/(2M_\pi)^2=q_1q_2/ 2M^2_\pi$ given by Eqs. (\ref{13}) and (\ref{14}) with
the integral representation
\ba
{\Phi}(t_\pi)&=& 6\int\limits_{0}^{1} dx \int\limits_{0}^{1} dy y\frac{4y^2x(1-x)}
    {1-4y^2x(1-x)t_\pi}=1+ t_\pi \frac{8}{15}...
\ea
In the limit $m^2_\sigma=4m_u^2+M_\pi^2 \to \infty$
we get the
 result  obtained in
 \cite{1979,75} 
\ba
\beta^{\rm  pion}(t)&=&
 \left[t_{\pi}-\frac{1}{4}\right]\frac{{\Phi}(t_\pi)}{3},
\ea
with the   chiral symmetry breaking given by Eq. (\ref{123}).

Thus,
the sum of contributions of all loops
   takes the form of the dynamical pion polarizability
\ba\label{tot-1}
\beta_{\pi^\pm}^{\rm NJL}(t) = \beta^{\rm quark}_{\pi^\pm}(t)+\beta^{\rm pion}_{\pi^\pm}(t)=
 \frac{m^2_\sigma-M_\pi^2}{m^2_\sigma-t}\left[\frac{10}{9}+
 \left(t_\pi -\frac{1}{4}\right)\frac{\Phi(t_\pi)}{3}\right]-\frac{1}{9}.
\ea
This  pion polarizability is in agreement with the results obtained in  the linear sigma model
 \cite{1977}
and the infinite sigma mass   limit of the nonlinear Chiral Lagrangians \cite{75}.

The Fig. \ref{fig2b} shows us all contributions taken separately
and  their sum (quarks + mesons) for $m^2_\sigma=4m_u^2+M_\pi^2$,   $m_u = 280$ MeV in terms of
$t_\pi=t/(2M_\pi)^2$.

We can see that there is the sigma pole effect of the  t-dependence
of the effective pion polarizability.
This  effect can explain different results of the different
experiments given in Introduction (\ref{p-1e}), (\ref{p-1ee}), and (\ref{p-1eee}).

 \begin{figure}[h]
\begin{center}
\includegraphics[width=25pc]{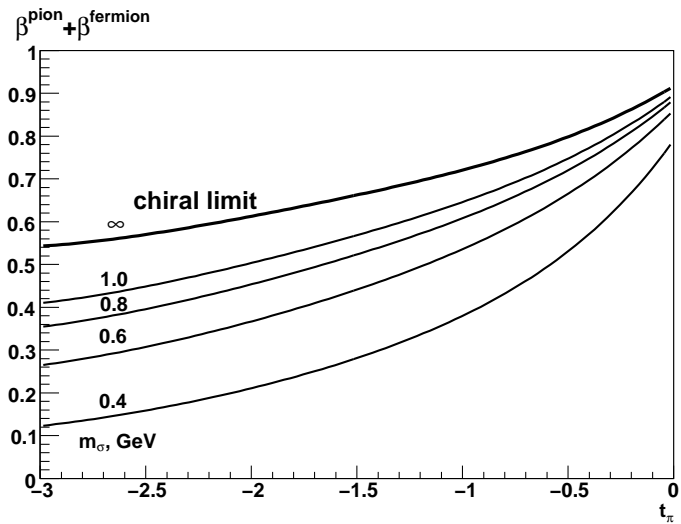}
\caption{\label{pic:beta}\small $\beta^{baryon(quark)}+\beta^{pion}$ for different values of $m_{\sigma}$.}
\end{center}
\end{figure}
 \begin{figure}[h]
\begin{center}
\includegraphics[width=25pc]{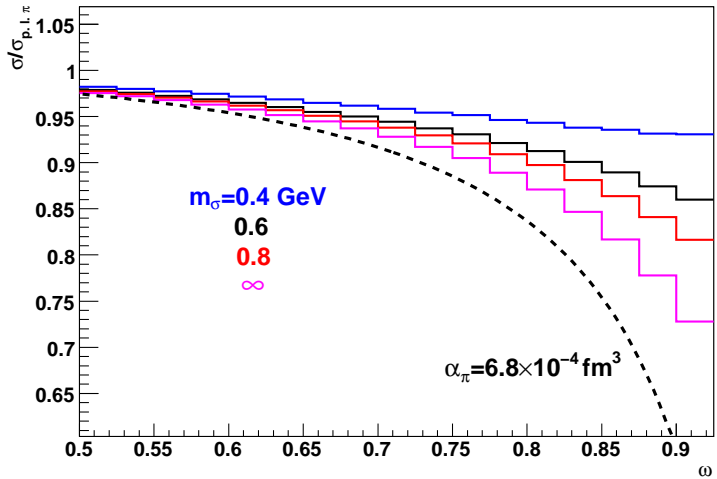}
\caption{\label{pic:omega}\small The ratio of the differential
 cross sections $d\sigma/d\omega$ for different values of $m_{\sigma}$.}
\end{center}
\end{figure}
 \begin{figure}[h]
\begin{center}
\includegraphics[width=25pc]{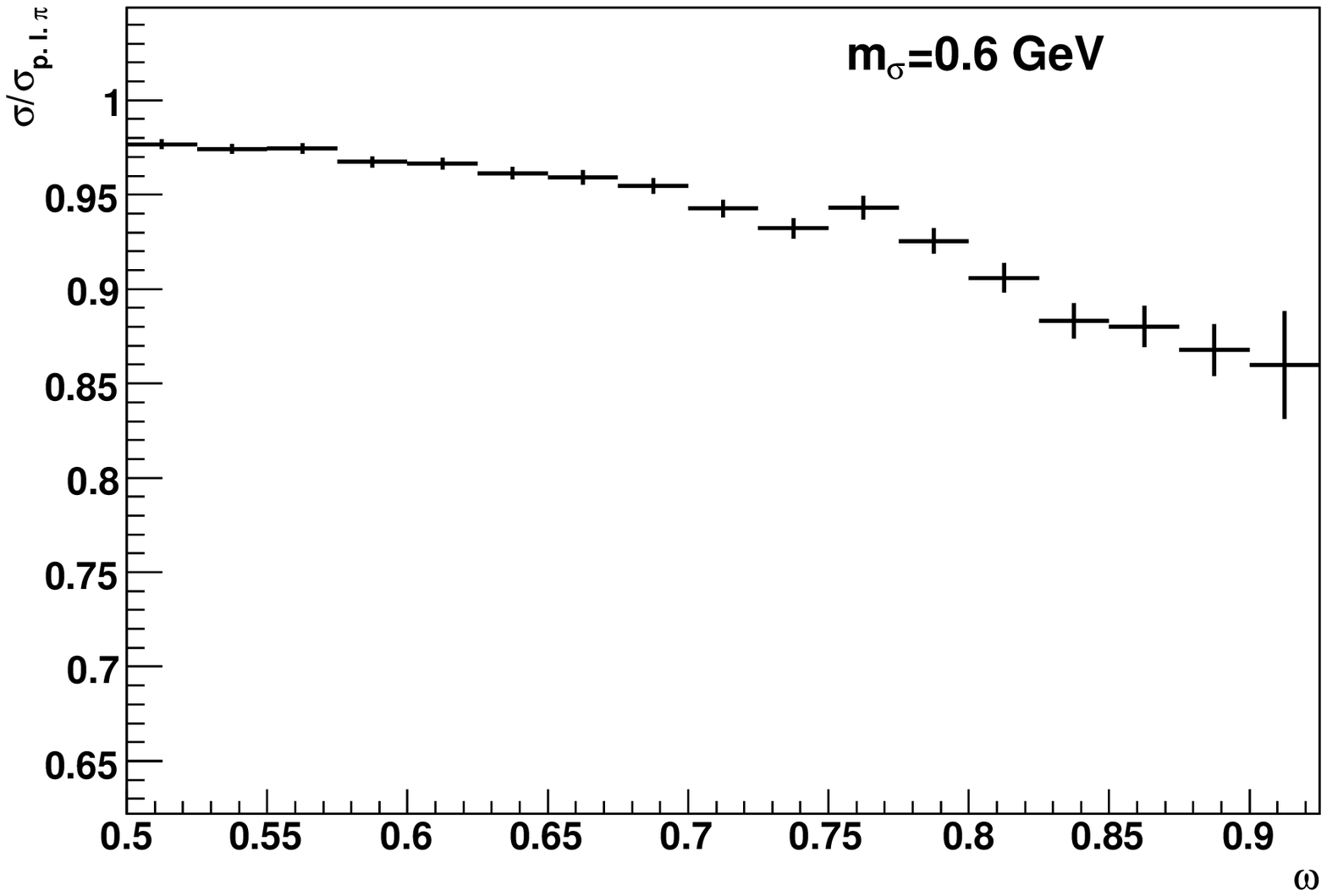}
\caption{\label{pic:omega2}\small The MC simulation for ratio of the differential
cross section $d\sigma/d\omega$ corresponding to 5 months of data taking ($m_{\sigma}=0.6~GeV$).}
\end{center}
\end{figure}

\section{Possibility of the experimental test of the prediction of NJL model at COMPASS}

 The COMPASS is the fixed target experiment in the secondary beam of Super
 Proton Synchrotron (SPS) at CERN. The purpose of this experiment is the study
 of hadron structure and hadron spectroscopy with high intensity muon and hadron beams.
 The COMPASS setup
provides unique conditions for  investigation of the process (1)
(\cite{compass},\cite{Moinester},\cite{sasha},\cite{moin}).
It has silicon detectors up- and downstream of the target  for the precise vertex
position reconstruction and for the measurement of the pion scattering angle,
an electromagnetic calorimeter for the photon 4-momentum reconstruction and
two magnetic spectrometers for the determination of the scattered pion momentum.
Hadron calorimeters and muon identification system
can be used for identification of secondary particles.

The kinematic range, covered by the COMPASS experiment approximately corresponds to
the parameter values range
$$0.5<\omega<0.95,~~~~~~M_\pi/3<p_t< 2M_\pi,~~~~s=(p_2+q_2)^2=
\frac{M^2_\pi\omega+p_t^2}{\omega(1-\omega)}<(3.75M_\pi)^2$$

$$-{2}~<~\frac{t}{(2m_\pi)^2}
<-\frac{1}{8}~.$$
 The Monte Carlo simulation,
based on the realistic description of the COMPASS detector using GEANT3 toolkit,
was performed to study the interaction of  190 GeV/c $\pi^{-}$ beam with 5 mm nickel target.
High intensity of the hadron beam (up to $2\times10^7$ pions per 10 s spill) and
 high capabilities of the trigger and DAQ system will allow to collect enough statistics of
 $\pi^{-}+A\to A+\pi^{-}+ \gamma$ events for precise measurement of pion polarizabilities.
 COMPASS will able to measure the pion polarizabilities  not only averaged
over some kinematic region (as it was done in Serpukhov experiment
\cite{Serp1} and MarkII \cite{MarkIICollab}) but also $\alpha_{\pi}$ dependencies on the kinematic variables.

The possibility to extract the mass of $\sigma$-meson from the behavior of the differential
cross section $d\sigma/d\omega$ was studied basing on the assumptions that $\alpha_{E}+\alpha_{H}=0$
and that baryon and pion loops contribute to $\alpha_{\pi^{\pm}}$ (see Fig \ref{pic:beta}). The ratio of
the differential cross section $d\sigma/d\omega$, predicted by NJL model, to the corresponding cross
section for point-like pion is presented in Fig. \ref{pic:omega}.  Fig. \ref{pic:omega2} shows the
result of the simulation for $m_{\sigma}=0.6~GeV$ for the case of $10^6$ events which corresponds to
 the total beam flux is $4\times10^{12}$ pions (approximately 5 months of running with beam intensity
 $2\times10^7$ pions per 10 s spill). The corresponding statistical error of the measurement of the mass
 of $\sigma$-meson is 25 MeV. For $m_{\sigma}=1.0 ~GeV$ statistical error increases to 90 MeV.

\section{Discussion and Conclusion}

The charge pion polarizability
was considered within the Nambu-Jona-Lazinio model.


%

In the difference with the earlier papers \cite{Volkov:1986zb,1985} on the NJL calculation of
pion polarizability, where only the quark loops were taken into account, here we calculated
the contributions of the meson loops.
However, in the case of the Compton effect, the noticeable contribution of the sigma pole
diagrams in Figs. 2,3 was
strongly concealed
by a set of diagrams with
internal sigma meson line in Fig. 4. 

At the region of  transfer $t\ll M^2_{\pi}$,
the prediction of the NJL model almost  coincides
within 5 -- 10\% of accuracy
with the QFT approach to Chiral Lagrangian \cite{1979}, where fermion loops
were taking into account. 

Thus, the NJL model result reveals
  the dominant role of the mean field approximation and the sigma pole diagram.
  This  sigma-pole diagram contains  the strong
t-dependence of the measurable effective polarizability
at the region of  transfer $t$ of the COMPASS experiment.
 This t-dependence
 can explain difference of two experimental results obtained in
 \cite{Serp1,9e} and by MARK II
 \cite{MarkIICollab} at the different transverse momentum.

If the experimental uncertainties is less then 5\% then pion polarizability
dependence on transverse momentum can be measured in the region of the variation
of the observable parameters.

\section{Acknowledgements}

The authors want to thank Drs. A.B. Arbuzov, S.B. Gerasimov, E.A. Kuraev, and M.A.
 Ivanov for fruitful discussions.

{}

\end{document}